\documentclass[epj]{svjour}
\usepackage{graphics}
\usepackage{rotating}
\usepackage{subfigure}
\begin{document}
\title{Hydrogen Spectroscopy with a Lamb-shift Polarimeter}
\subtitle{An Alternative Approach Towards Anti-Hydrogen Spectroscopy Experiments}
\author{M.~P.~Westig\inst{1,2} \and R.~Engels\inst{2} \and K.~Grigoryev\inst{2,3}
\and M.~Mikirtytchiants\inst{2,3} \and F.~Rathmann\inst{2} \and 
H.~Paetz gen.~Schieck\inst{4} \and G.~Schug\inst{2} \and A.~Vasilyev\inst{3} 
\and H.~Str\"oher\inst{2}
}                     
\institute{I.~Physikalisches Institut, Universit\"at zu K\"oln, 50937 K\"oln, 
Germany \and
Institut f\"ur Kernphysik and J\"ulich Center for Hadron Physics, 
Forschungszentrum J\"ulich GmbH, 52425 J\"ulich, Germany \and
Petersburg Nuclear Physics Institute, 188300 Gatchina, Russia \and
Institut f\"ur Kernphysik, Universit\"at zu K\"oln, 50937 K\"oln, Germany}
\date{Received: date / Revised version: date}
%
\abstract{A Lamb-shift polarimeter, which has been built for a fast
determination of the polarization of protons and deuterons of an
atomic-beam source and which is frequently used in the ANKE
experiment at COSY-J\"ulich, is shown to be an excellent device for
atomic-spectroscopy measurements of metastable hydrogen isotopes. It
is demonstrated that magnetic and electric dipole transitions in
hydrogen can be measured as a function of the external magnetic
field, giving access to the full Breit-Rabi diagram for the
$2^2S_{1/2}$ and the $2^2P_{1/2}$ states. This will allow the study
of hyperfine structure, $g$ factors and the classical Lamb shift.
Although the data are not yet competitive with state-of-the-art
measurements, the potential of the method is enormous, including a
possible application to anti-hydrogen spectroscopy.
} 
\maketitle
The spectrum of atomic hydrogen (H), unravelled with ever increasing
precision, has led to fundamental understanding of the
underlying structure and dynamics of the simplest atom. Today, both
experiment (e.g., two-photon laser
spectroscopy~\cite{udem97,huber98}) and theory~\cite{pachucki96} have
achieved an impressive level of accuracy in the determination of the
energy levels. Additional details and further discoveries may come
from either advancing existing techniques or by applying new
methods.

\begin{figure}[htb]
\begin{minipage}[c]{\columnwidth}
\includegraphics[width=\columnwidth]{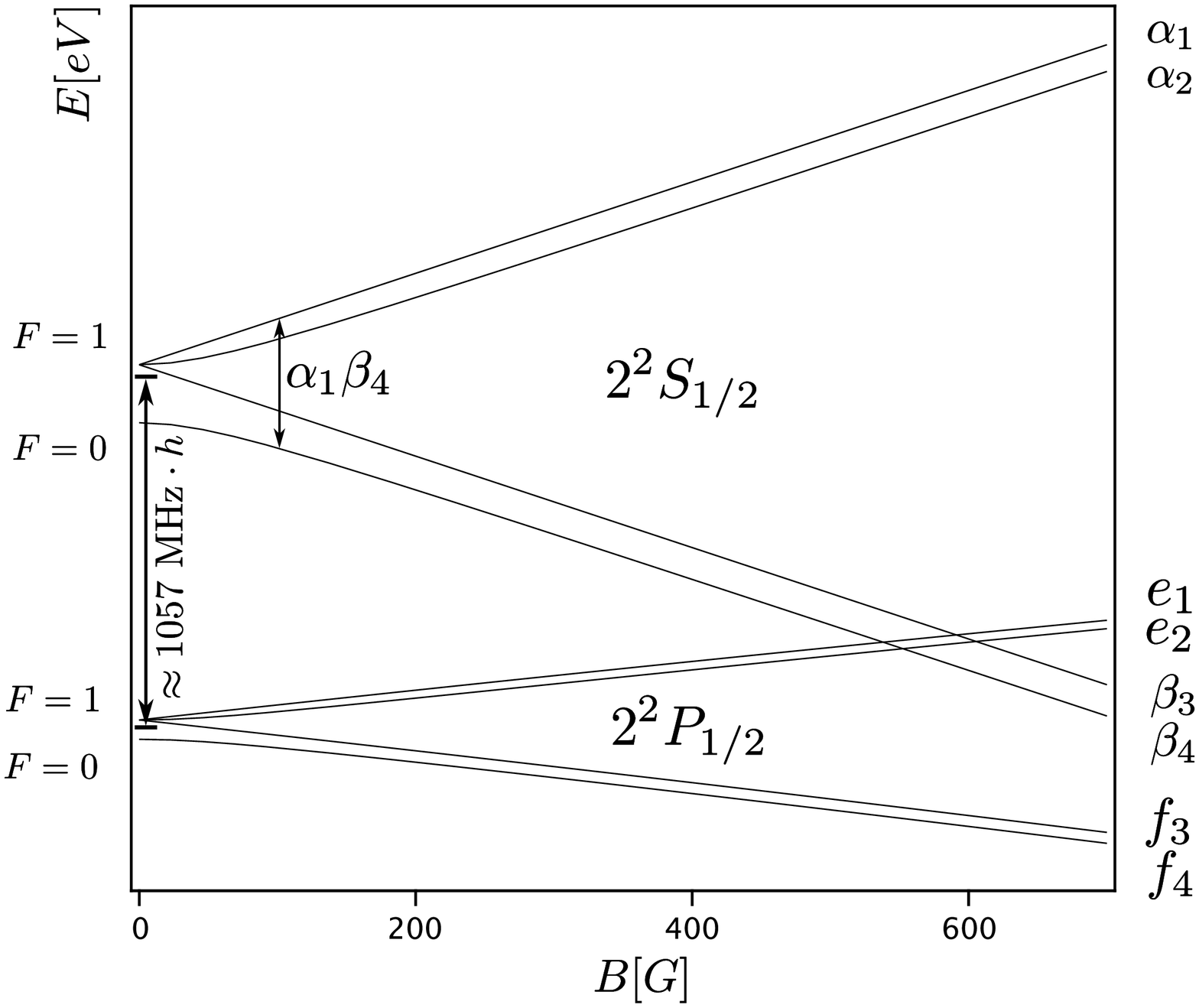}
\end{minipage}
\vspace{3mm}\\
\begin{minipage}[c]{\columnwidth}
\begin{tabular}{l|l|l}
\multicolumn{1}{c}{State}&\multicolumn{1}{|c|}{Zeeman Region}&\multicolumn{1}{c}{Paschen-Back Region}\\
\hline
$\alpha_1$&$|F=1,m_{F}=1\rangle$&$|m_{j}=1/2,m_{I}=1/2\rangle$\\
$\alpha_2$&$|1,0\rangle$&$f^{+}|1/2,-1/2\rangle + f^{-}|-1/2,1/2\rangle$\\
$\beta_3$&$|1,-1\rangle$&$|-1/2,-1/2\rangle$\\
$\beta_4$&$|0,0\rangle$&$f^{-}|1/2,-1/2\rangle - f^{+}|-1/2,1/2\rangle$\\
\end{tabular}
\caption{Breit-Rabi diagram of H for the first excited state with
$j=1/2$~\cite{moskovkin06,moskovkin07}. The classical Lamb shift of
1057~MHz and a magnetic dipole transition from the state $\alpha_1$
into $\beta_4$ are indicated. The table indicates the definitions for
the Zeeman and the Paschen-Back regions for the $\alpha$- and
$\beta$-states, where $f^{\pm}=1/\sqrt{2}\sqrt{1\pm a(B)}$ are the
critical field parameters describing the transition between them and
$a(B)=B/\left(B_{C}\sqrt{1+(B/B_{C})^2}\right)$ with the critical field 
strength $B_{C}=\Delta E_{\mathrm{HFS}}/2\mu_B$ where $\Delta E_{\mathrm{HFS}}$ denotes the
energy separation of the hyperfine states of the $2^2S_{1/2}$ 
state and $\mu_B$ is the Bohr
magneton. Corresponding definitions apply for the state $2^2P_{1/2}$.}
\label{fig1}
\end{minipage}
\end{figure}

A possible new approach to performing spectroscopy experiments on
hydrogen isotopes is based on a Lamb-shift polarimeter which
would allow measurements of the full Breit-Rabi diagram of the
first excited state of H with $j=1/2$ (see Fig.~\ref{fig1}).
Precision data currently exist only for weak magnetic fields of a few
Gauss for the ground state~\cite{schmidtkaler95}. For the metastable
state, the crossing of the $2^2S_{1/2}$ and $2^2P_{1/2}$ states
($\beta-\mathrm{e}$ crossing of Fig.~\ref{fig1}) at field strengths
of about 570~G was investigated long ago~\cite{lamb51}. In addition
to testing the recent relativistic theory of the Zeeman splitting of
the first excited state~\cite{moskovkin06,moskovkin07}, systematic
Breit-Rabi diagram measurements can also be used to 
extract $g_j(\mathrm{H},2^2S_{1/2})$,
$g_j(\mathrm{H},2^2P_{1/2})$ and the nuclear $g$ factor. The most recent
determination of a hydrogen $g_j$ factor for the ground state
was performed by Tiedeman and Robinson~\cite{tiedeman77}. Mass 
independent terms of quantum electrodynamics 
could be tested to a high degree and a confirmation of the
$\alpha^3$ radiative correction has been achieved. In order 
to test, e.g., bound-state quantum electrodynamic corrections of the order
of $\alpha/\pi$ for the $g_j$ factor, 
a precision of $1$~ppb and better is needed
for the hydrogen atom in the ground state \cite{beier00}.
An even better precision would be needed for the 
first excited state. By putting the
hydrogen atom in an external magnetic field, the independence of the
hyperfine structure to such conditions can be tested through a
combination of measurements of transitions between $2^2S_{1/2}$ and
$2^2P_{1/2}$ states (Table~\ref{tab1}).

In 1940, Kusch et al.\ performed spectroscopy experiments on alkaline
atoms in a magnetic field with an improved molecular beam resonance
method for atoms and determined the hyperfine structure 
and nuclear moments from
measurements of the Breit-Rabi diagram~\cite{kusch40}. 
Though Rabi realized that
this method should also be applicable to excited atomic
states \cite{rabi52}, it seems never to have been exploited before.
%
%
%
%
\begin{figure}[t!]
\centering
\setcounter{subfigure}{0}
\subfigure[]
{\label{fig2a}\includegraphics[width=0.45\columnwidth]{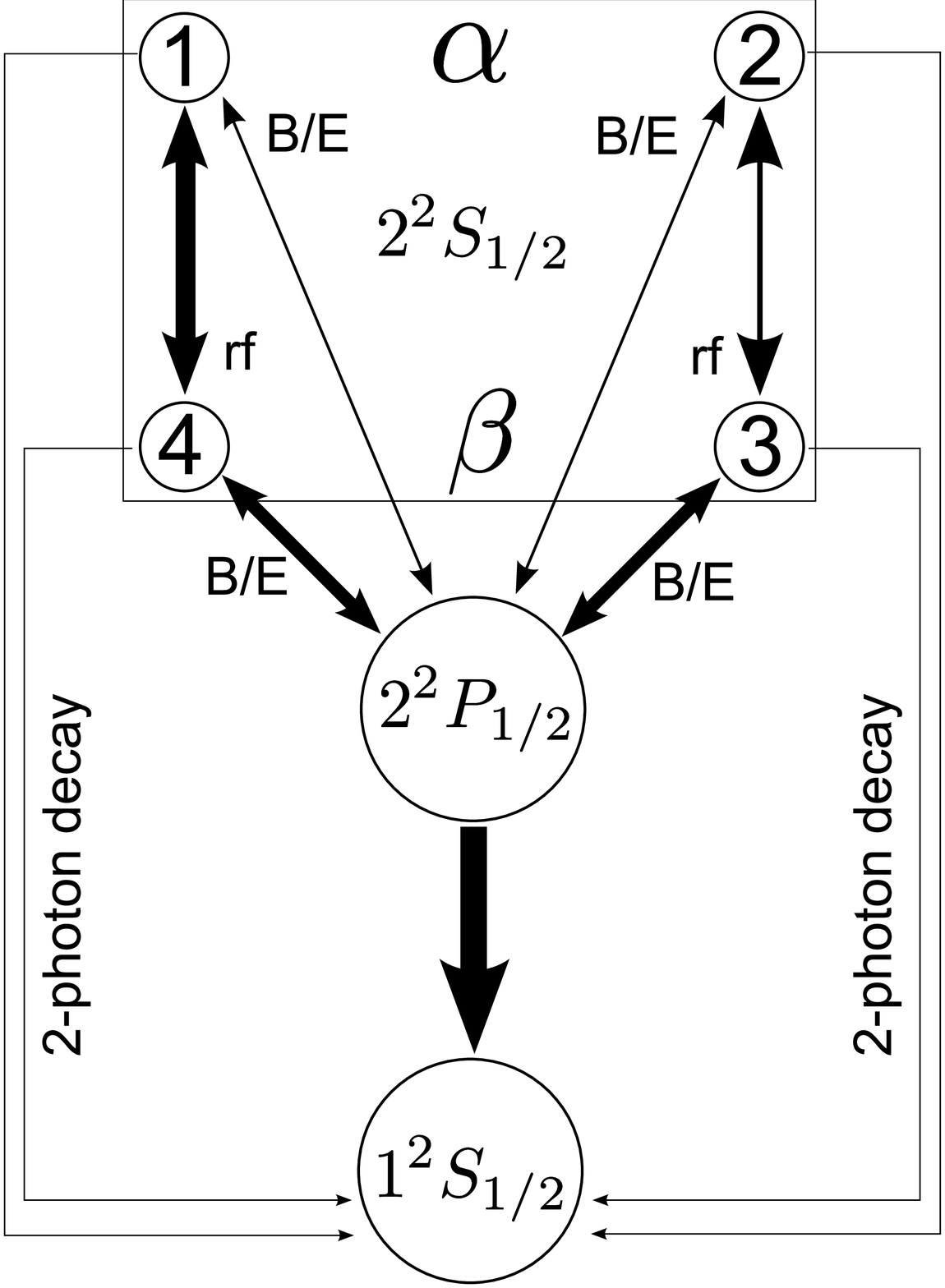}}
\subfigure[]
{\label{fig2b}\includegraphics[width=\columnwidth]{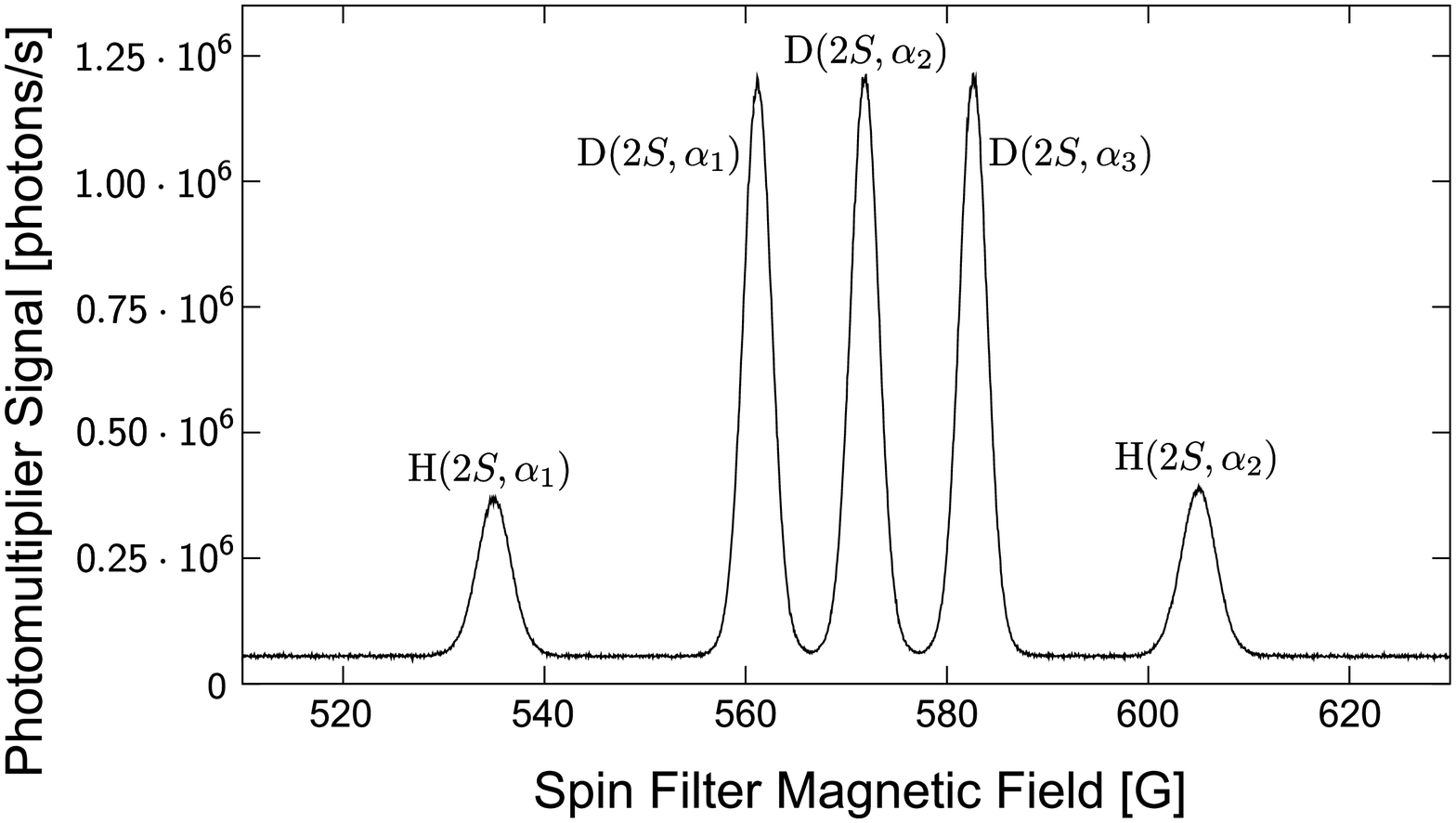}}
\caption{(a) Principle of the spin-filtering technique for metastable hydrogen. 
In the absence of magnetic fields, $2^2S_{1/2}$ states can decay to the
$1^2S_{1/2}$ state by 2-photon emission.
However, the simultaneous interaction with 
a magnetic field $B$, an electric field $E$ and 
a radio-frequency ({\emph {rf}}) field, allows one to select, 
e.g., state $\alpha_1$: by tuning the 
magnetic field to 535~G, state $\beta_4$ intersects with
$e_1$ while $\beta_3$ is very close to the 
short-lived $2^2P_{1/2}$ states. An electric field mixes
$\beta_3$ and $\beta_4$ with the $2^2P_{1/2}$ state with subsequent 
decay to the ground state. The radio-frequency field strongly 
couples $\alpha_1$ and $\beta_4$ (thick arrow), while it weakly
connects $\alpha_2$ and $\beta_3$ (thin arrow). This leads to an
emptying of $\alpha_2$ by means of $\beta_3$ while
$\alpha_1$ stays in oscillation with $\beta_4$.
Turning off the radio-frequency field, residual atoms in the state
$\beta_3$ and $\beta_4$ are quenched into the ground 
state and only atoms in
state $\alpha_1$ are left in the beam. The same principle 
applies to the $\alpha_2$ state and analogously to the deuterium atom.
(b) Lyman-$\alpha$ spectrum for metastable hydrogen and 
deuterium as a function of the magnetic field in the
spin filter after optimization of intensity and state separation.
The spectrum has been obtained by a photomultiplier on top of the
transition unit (see Fig.~\ref{fig3a}).
For hydrogen, states $\alpha_1$ and $\alpha_2$ can be selected by
tuning the magnetic field to 535~G and 605~G, respectively.
For deuterium, the field strengths are $B=$ 565~G ($\alpha_1$),
575~G ($\alpha_2$) and 585~G ($\alpha_3$).}
\label{fig2}
\end{figure}
%
%
The measurements of Lundeen and Pipkin 
\cite{lundeen81,lundeen86} provide the best direct 
determinations of the classical Lamb shift.
Their experiment used transitions between the hyperfine structure of the
$2^2S_{1/2}$ and $2^2P_{1/2}$ states. In contrast, we present an enhanced
experimental method to measure transitions between single Zeeman components
of the hyperfine structure with a spin-filtering technique. 
Besides the hyperfine structure and the 
classical Lamb shift, the measurement of these transitions
will allow to determine the Breit-Rabi diagram 
of the $2^2S_{1/2}$ and the $2^2P_{1/2}$ state
as is shown in this paper.

At the Forschungszentrum J\"ulich, we have built an atomic-beam
source to produce beams of polarized hydrogen and
deuterium~\cite{grigoryev07} for use as internal targets in experiments
at the cooler storage ring COSY~\cite{maier01}. The atomic-beam source has been
complemented by a Lamb-shift polarimeter~\cite{engels03,engels05} for the fast on-line
determination of the target polarization. When not required for this
purpose, we have started to use the 
Lamb-shift polarimeter - atomic-beam source combination to study
atomic hydrogen spectroscopy in a dedicated laboratory setup. The
results of a proof-of-principle experiment described here demonstrate
that a metastable beam of atoms produced in a well-defined Zeeman
state ($\alpha_1$ or $\alpha_2$ of 
Fig.~\ref{fig1}, see also Fig.~\ref{fig2}) by the 
Lamb-shift polarimeter, will
be an excellent tool for precision hydrogen spectroscopy.

\begin{figure}[t]
\setcounter{subfigure}{0} 
\subfigure[]
{\label{fig3a}\includegraphics[width=\columnwidth]{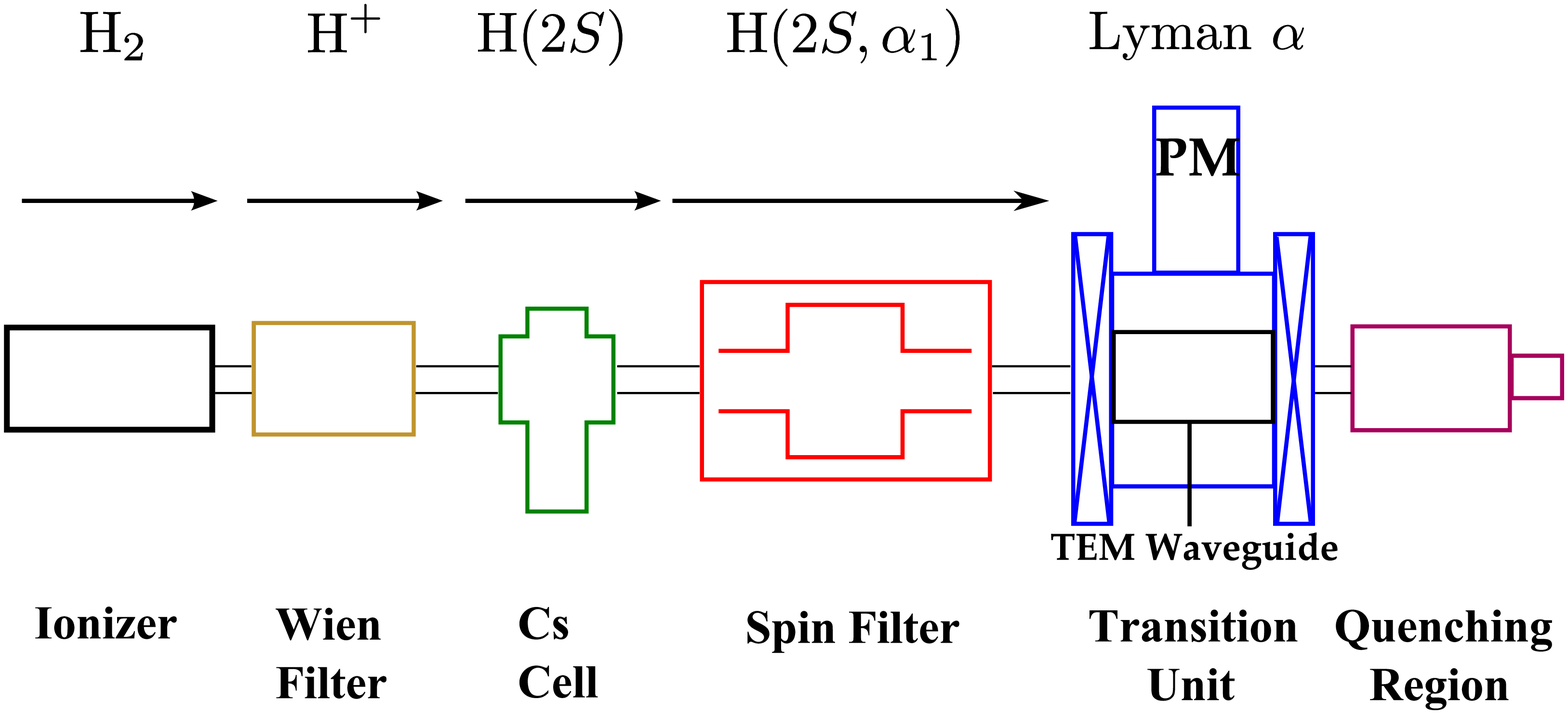}}
\subfigure[]
{\label{fig3b}\includegraphics[width=\columnwidth]{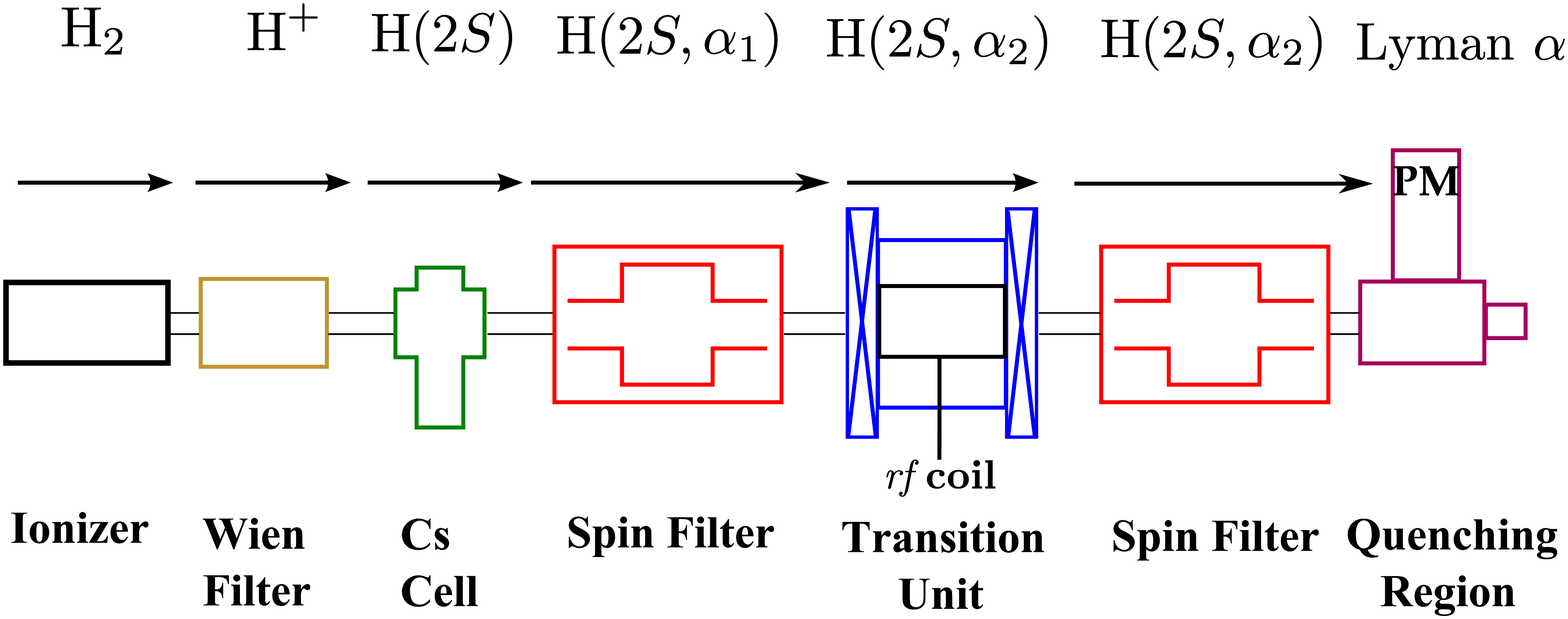}}
\caption{(color on-line) (a) Lamb-shift polarimeter 
setup used for the measurements of electric 
dipole transitions with atoms in the $\alpha_1$ state. 
(b) Shows the changes necessary
for a possible future experiment to measure magnetic 
dipole transitions, where atoms in the
$\alpha_1$ state undergo transitions into the $\alpha_2$ state. 
The position of the photomultiplier will change for
measurements of electric dipole and magnetic dipole 
transitions. The transition unit is surrounded by a
pair of Helmholtz coils. A future upgrade would replace the
ionizer, the Wien filter and the cesium cell by a cold excited
atomic beam (see text).} \label{fig3}
\end{figure}

We start by briefly recalling the operating principles of the 
atomic-beam source
and the Lamb-shift polarimeter. 
The atomic-beam source which might be used in a future setup 
provides a cold atomic beam emerging from a
dissociator and a cooled nozzle. Atoms with a definite total angular
momentum are filtered by a Stern-Gerlach assembly of sextupole
magnets and specific hyperfine states are then selectively populated by
radio-frequency units. The Lamb-shift polarimeter (see Fig.~\ref{fig3}) 
ionizes the atomic beam in a strong magnetic
field to preserve the nuclear spin. After the metastable state is
produced in a cesium cell, a spin filter~\cite{mckibben68}, a
quenching region with a static electric field and a photomultiplier,
sensitive to 121~nm Lyman-$\alpha$ radiation, is used to determine
the beam polarization through the relative occupation of the hyperfine
states~\cite{engels03} (see Fig.~\ref{fig2}). In the experiment described below the 
Lamb-shift polarimeter is used to produce metastable atoms 
in one single Zeeman state of the hyperfine structure 
($\alpha_1$, $\alpha_2$ and together with a Sona transition 
$\beta_3$~\cite{sona67}, see Fig.~\ref{fig1}).
%
%
%
%

The experimental setup is shown schematically in \linebreak 
Fig.~\ref{fig3},
where the upper part presents the situation for measuring electric
dipole transitions. H$_2$ molecules are first ionized inside
an electron-impact ionizer. After passing through a Wien filter,
which only transmits protons, the ion beam enters a cesium cell in
which metastable hydrogen atoms in the $2^2S_{1/2}$ state are
produced by the charge-exchange reaction $\textrm{H}^+ + \textrm{Cs}
\rightarrow \textrm{H}(2^2S_{1/2}) + \textrm{Cs}^+$. The beam energy
of 1~keV is a compromise between working at the maximum cross 
section for the process~\cite{pradel74} and minimizing losses due 
to beam divergence. 
Subsequently, one of the Zeeman states of the hyperfine structure 
($\alpha_1$ or $\alpha_2$) is selected in a spin filter 
(see Fig.~\ref{fig2}). Finally, 
these atoms enter the transition unit with a
Lecher TEM (transverse electromagnetic) waveguide, which produces an
electric field $\vec{E}$ parallel to the beam axis.
%
%
\begin{figure}[t]
\subfigure{\label{fig4a}\includegraphics[width=\columnwidth]{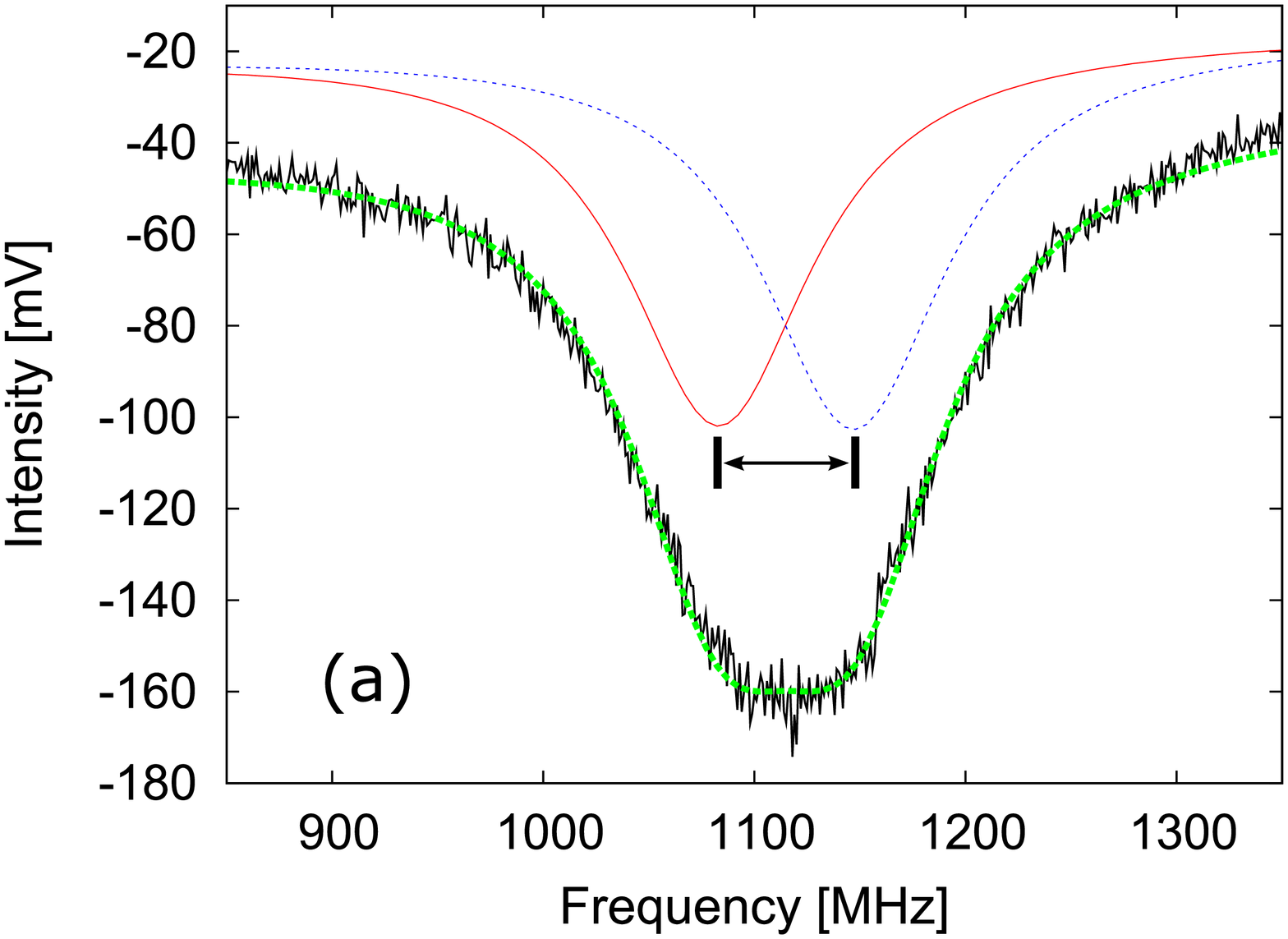}}\\
\subfigure{\label{fig4b}\includegraphics[width=\columnwidth]{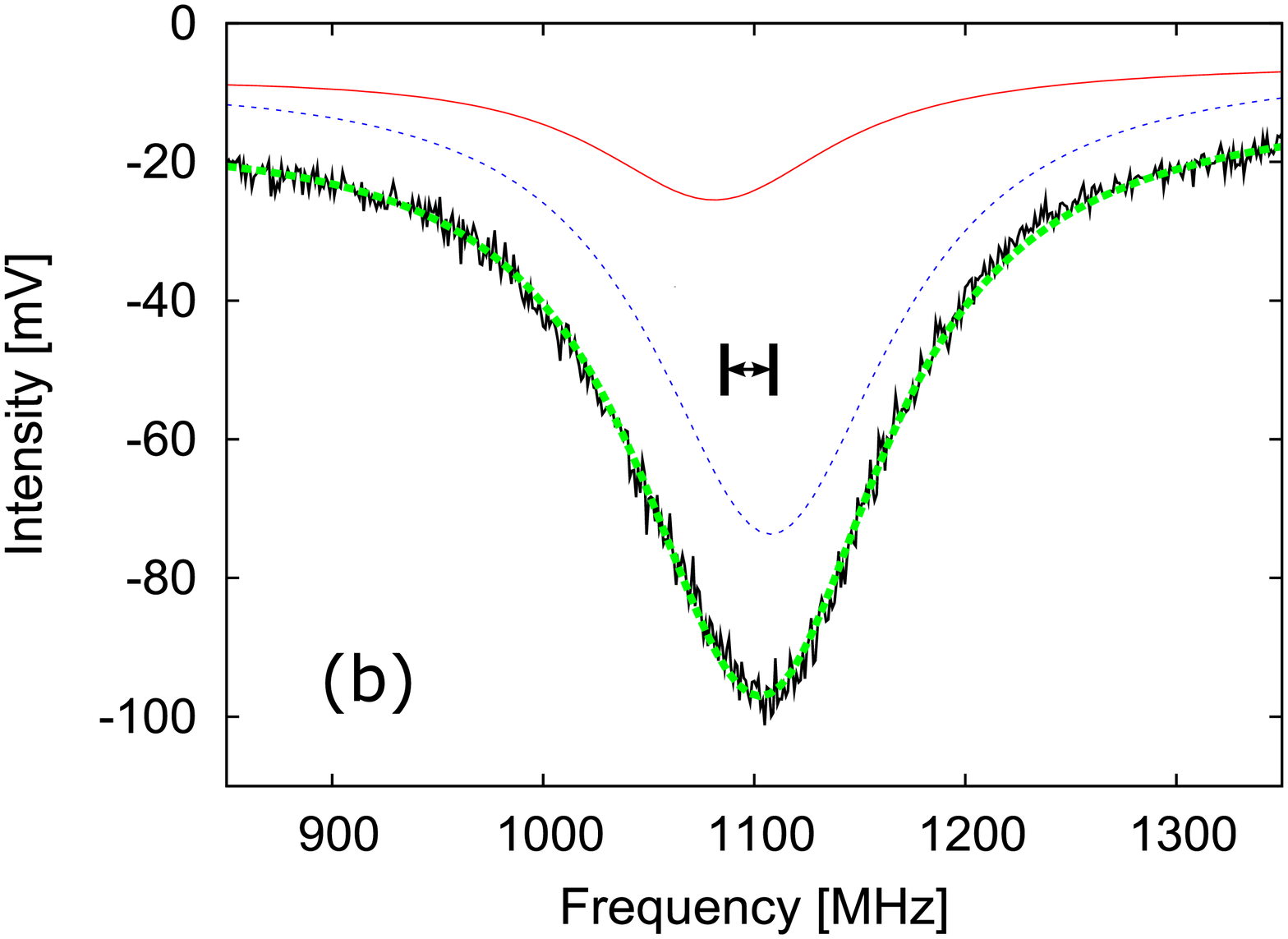}}
\caption{(color on-line) (a) Result for electric dipole transitions
($\sigma$ transitions) at small external magnetic field strength ($B
\approx 0$~G) for a beam of atoms in the $\alpha_1$ state. 
The thin black solid line is the measured spectrum while
the thick green dashed line is the fit to the data. This is the sum
of two single fits shown as thin red solid and dashed blue lines
corresponding to the transitions $\alpha_1 \rightarrow e_2, f_4$. 
The spacing between the fitted minima corresponds to the $2^2P_{1/2}$
hyperfine structure, indicated by the black arrow. (b) Electric
dipole transitions ($\sigma$ transitions) in an
external magnetic field of strength $B=26.8(5)$~G for a beam of
atoms in the $\alpha_2$ state undergoing transitions into the states
$e_1$ and $f_3$. With increasing magnetic field strength, the transition 
probability into the state $e_1$ clearly decreases (thin red solid line).
The black arrow indicates the spacing between the Zeeman states
$e_1$ and $f_3$ (dashed blue line) of the $2^2P_{1/2}$ hyperfine 
structure component with total angular momentum quantum number $F=1$.}
\end{figure}
%
%

Figure~\ref{fig3b} shows the setup for the measurement of magnetic
dipole transitions. Here a second spin filter is added between
the transition unit and the quenching region, the TEM waveguide is
replaced by a radio-frequency coil, and the photomultiplier is moved 
to the quenching region.

First data were obtained by measuring electric dipole transitions
($\sigma$ transitions, $\vec{B} \perp \vec{E}$) to the
$2^2P_{1/2}$ state inside the TEM waveguide. Because of their short
lifetime ($10^{-9}$~s) the atoms will immediately decay to the ground
state, emitting Lyman-$\alpha$ photons. These are detected through a
hole in the TEM waveguide by a photomultiplier mounted on top of the
transition unit.
In Fig.~\ref{fig4a}, the result of such a measurement
is shown where the external magnetic field was close to zero. The
intensity of Lyman-$\alpha$ photons is plotted as a function of the
TEM radio frequency. The spectrum, obtained in about 10~min measuring
time, contains $\approx 10^7$ photons. 
The peak broadening is due to
the superposition of two transitions ($\alpha_1 \rightarrow e_2,f_4$,
Fig.~\ref{fig1}) and a two-Lorentzian fit results in the shapes shown
in Fig.~\ref{fig4a}. The frequency difference corresponds to the
$2^2P_{1/2}$ hyperfine splitting $f_{\mathrm{HFS}}(2^2P_{1/2})$ 
for which we obtain $60 (2)$~MHz. Although this agrees
with theory (59221.2~kHz~\cite{moskovkin07}), it has a large
error. Our result for the classical Lamb shift measured in the same
experiment, 1057(1)~MHz agrees with theory and recent
experiments~\cite{pachucki96}, but the error is two orders of
magnitude too large to be competitive. Possible ways to reduce these
errors significantly are discussed below.

In Fig.~\ref{fig4b} we show a measurement of atoms in the
$\alpha_2$ state undergoing transitions into the states $e_1$ and 
$f_3$ in an external magnetic field of strength $B=26.8(5)$~G. 
In this case the measured spacing between the states $e_1$ and $f_3$
is $27(2)$~MHz compared to the expected value of
$24.855$~MHz \cite{moskovkin06,moskovkin07}.
We have to verify this value with the hardware improvements of our
experiment described at the end of this text.   
Measurements of this kind will allow one to determine the whole 
Breit-Rabi diagram of the $2^2S_{1/2}$ and the $2^2P_{1/2}$ states.

We have also performed measurements with a magnetic field close to
zero, oriented parallel to the TEM radio-frequency field 
(electric dipole transitions, $\pi$
transitions, $\vec{B} \parallel \vec{E}$).
In this case, selection
rules led to the possible transitions $\alpha_1\rightarrow e_1$ and
$\alpha_2\rightarrow e_2, f_4$.
%
%
\begin{table}[t]
\caption{Transition-frequency combinations in hydrogen
and deuterium to determine magnetic-field-independent values of the
hyperfine structure. Here, e.g., $\alpha_1\beta_4$ denotes the
transition $\alpha_1 \to \beta_4$. (See also Fig.~\ref{fig1}).}
\label{tab1}
\centering
\begin{tabular}{ll}
\hline\noalign{\smallskip}
\multicolumn{2}{c}{{\bf{Hydrogen}}}\\
$f_{\mathrm{HFS}}(2^2S_{1/2})$&$\alpha_1\beta_4-\alpha_2\beta_3$\\
&$\alpha_1\alpha_2+\beta_3\beta_4$\\
$f_{\mathrm{HFS}}(2^2P_{1/2})$&$\alpha_1f_4-\alpha_2f_3-\alpha_1e_1+\alpha_2e_2$\\
\\
\multicolumn{2}{c}{{\bf{Deuterium}}}\\
$f_{\mathrm{HFS}}(2^2S_{1/2})$&$\alpha_1\beta_5-\alpha_3\beta_4+\alpha_1\beta_6-\alpha_2\beta_4$\\
&$\alpha_1\alpha_2+\alpha_1\beta_6+\beta_4\beta_5-\alpha_3\beta_4$\\
$f_{\mathrm{HFS}}(2^2P_{1/2})$&$\alpha_3e_3+\alpha_2f_5-\alpha_2e_1-2\alpha_3f_4$\\&$-\alpha_1e_1+
\alpha_3e_2+\alpha_1f_6$\\
\noalign{\smallskip}\hline
\end{tabular}
\end{table}
%
%
As in the case shown in Fig.~\ref{fig4b}, for atoms in the state $\alpha_2$ a
decrease in the probability of a transition into state $f_4$ 
is observed if $B$ exceeds the critical field value $\Delta E_{\mathrm{HFS}}/2\mu_B$,
i.e.,~$a(B)\rightarrow 1$ (see Fig.~\ref{fig1}), which is the
Paschen-Back effect.
These atoms only undergo transitions into state
$e_2$ and, as a result, a narrower line shape is obtained than that
found in Fig.~\ref{fig4a}.

When measuring magnetic dipole transitions, the setup has to be
modified, as shown in Fig.~\ref{fig3b}. Behind the first spin filter,
atoms in one Zeeman state of the hyperfine structure 
enter the spectroscopy region where magnetic dipole transitions are
induced inside a radio-frequency coil. A second spin filter is subsequently
used to check the occurrence of a definite transition between two
Zeeman components of the $2^2S_{1/2}$ hyperfine structure. 
It is, therefore, tuned to transmit only one particular state.

The photomultiplier on top of the
quenching region monitors the occurrence of such a transition during
Stark-effect quenching of the residual metastable atomic beam. This
scheme is actually very similar to the one suggested by
Rabi~\cite{rabi52}.

Since the electric field is parallel to the beam axis, the wave vector is
perpendicular and the longitudinal Doppler effect is suppressed in
first order. A signal generator supported by a radio-frequency amplifier
was connected to the TEM waveguide where a power meter was used to
keep the radio-frequency power level constant. The spectroscopy region
itself consists of a commercial stainless steel chamber with a pair
of magnetic field coils surrounding the chamber. The spectroscopy
region was not shielded against external magnetic fields in these
initial measurements.

Although proof-of-principle measurements with the \linebreak Lamb-shift 
polarimeter have been
achieved, further \linebreak progress has to be made to ensure that this is a
precision spectroscopic tool. Some components of the current setup shown
in Fig.~\ref{fig3} must be modified. The stainless steel vacuum
chamber in the spectroscopy region should be replaced by an aluminum
chamber covered by a $\mu$-metal shielding in order to decrease the
influence of external electromagnetic fields on the resonance shape
and frequency. Since $\Delta B \approx 0.5$~G, the uncertainty in 
the frequency is about 1~MHz and an improved set of magnetic field
coils should be implemented in order to reach 
$\Delta B/B \approx 10^{-4}$. The variations
in the controlled radio-frequency power ($\approx 3\%$) lead to an
uncertainty of approximately $100$~kHz. With state-of-the-art
hardware, the power can be controlled on a $10^{-5}$ level.

In order to extract precise resonance frequencies from our measured
spectra, a detailed understanding of the shape is needed. The largest
effect arises from the uncertainty principle. Because of the short
interaction time of the atoms with the radio-frequency field, a resonance
broadening of approximately 45~MHz occurs.
For electric dipole 
transitions, the convolution with the natural resonance 
width results in a line width of approximately $110$~MHz, while 
for magnetic dipole transitions a value of $45$~MHz
is expected.

The atomic beam has a fixed velocity but, due to scattering and
electron capture in the cesium cell, effects of energy straggling of
about 1\% have to be taken into account. For measurements in magnetic
fields $\leq 100$~G errors for a single measurement of $10$~kHz for
the $2^2S_{1/2}$ state and $100$~kHz for the $2^2P_{1/2}$ state are
to be expected with this setup. Because the $g_j$ factor dominates
the slope of the Breit-Rabi diagram, it can be identified with a precision of
$10^{-5}$ for $2^2S_{1/2}$ and $10^{-4}$ for $2^2P_{1/2}$,
respectively. Although such a measurement will allow a precise
determination of the Breit-Rabi diagram, the $g_j$ factor could only be used to
check for relativistic effects. Hence an increase in the magnetic
field strength up to $1$~T is necessary to improve the $g_j$-factor
determination to a precision of $10^{-7}$ for $2^2S_{1/2}$ and
$10^{-6}$ for $2^2P_{1/2}$. 

This level of precision is not sufficient
to test bound-state quantum electrodynamic 
effects of the order of $\alpha/\pi$ for the
$2^2S_{1/2}$ state. To further improve the experimental setup, the
possibility of using the separated-oscillatory-field 
method~\cite{lundeen86} will be considered in the design of the new
spectroscopy region. Together with a reasonable enlargement of the
interaction region, this will decrease the uncertainty of the
measurement of magnetic dipole transitions by two orders of magnitude.

As a final step in the improvement, the atomic-beam source 
could be employed as the source of hydrogen atoms, 
which are excited, e.g., by electron impact.
The reduction of the atomic-beam velocity by more than two orders of 
magnitude would decrease the error for a single measurement of 
magnetic dipole transitions by the same amount.
By rotating the wave vector, the second order Doppler effect 
can also be cancelled.
The transverse Doppler effect, which increases the resonance
frequency, can be estimated by changing the beam energy. The motional
Stark effect has to be considered since this also shifts the energy
levels.

%
%
\begin{table}[t!]
\caption{Estimate of the projected precision for the
$g_j$ factor and the hyperfine structure 
that might be achieved during the further
development of the experiment. At this stage of development 
the Lamb-shift polarimeter will be used
together with a separated-oscillatory-field (SOF) setup and 
with a thermal beam emerging from an atomic-beam source (ABS).}
\label{tab2}
\centering
\begin{tabular}{ccccc}
\hline\noalign{\smallskip}
 & \multicolumn{2}{c}{$2^2S_{1/2}$} & \multicolumn{2}{c}{$2^2P_{1/2}$}\\
 & $g_j$ factor&HFS & $g_j$ factor&HFS\\
SOF & $10^{-7}$ & $100$~Hz & $10^{-6}$ & $1$~kHz\\
ABS & $<10^{-8}$ & $10-1$~Hz & - & -\\
\noalign{\smallskip}\hline
\end{tabular}
\end{table}
%
%

Measurements can be performed for different magnetic field settings
to obtain multiple field-independent values of the $2^2S_{1/2}$ and
$2^2P_{1/2}$ hyperfine structure which can be averaged 
to increase the precision.
Although the measurement of $f_{\mathrm{HFS}}(2^2P_{1/2})$ is
lifetime-limited, high-precision values of
$f_{\mathrm{HFS}}(2^2S_{1/2})$ are achievable. The classical Lamb shift
can also be extracted from the data with a precision $<10$~kHz. 
After implementation of all the discussed improvements, we
expect that measurements of $g_j$ factors and the hyperfine structure 
with the accuracies given in Table~\ref{tab2} can be achieved.

A very exciting prospect of the method discussed here is that
it could also be applied to beams of anti-hydrogen. 
Although a similar proposal has been put forward in \cite{meshkov98},
the method demonstrated in this paper provides additional useful 
features for such a measurement. During the
recombination of an antiproton with a positron, up to 30\% of the
anti-H atoms produced end up in the $2^2S_{1/2}$ state. Due to the
detection efficiency of $10^{-3}$, in order to detect 100 photons in
the photomultiplier, $3\times 10^6$ anti-H atoms are necessary.
Modifications of the quenching region should increase the detection
efficiency by two orders of magnitude, in which case 30000 atoms
would be sufficient to obtain the $2^2S_{1/2}$ hyperfine structure 
with a precision
of $140$~kHz. Reducing the beam energy down to 1~meV would improve
the precision to 5~kHz because the Doppler effect is smaller in this
case.

In summary, we have shown that a Lamb-shift polarimeter can be used for atomic
spectroscopy. The Lamb-shift polarimeter measurements of hyperfine 
structure transitions in atomic
hydrogen have not yet reached the level of accuracy that other
experiments have achieved for example for the $2^2S_{1/2}$ 
state~\cite{kolachevsky09,kolachevsky04,rothery00,heberle56}. 
However, the method will allow the determination of the $2^2P_{1/2}$ 
hyperfine structure with
up to two orders of magnitude higher precision than most recent
results~\cite{lundeen75}. In addition, this method is also applicable
to deuterium and might be used in the future for anti-hydrogen.

We are grateful for discussions with Th.~Udem, \linebreak  D.~L.~Moskovkin and
D.~A.~Glazov. We thank other members of IKP-2 of Forschungszentrum 
J\"ulich for their help during
the measurements and C.~Wilkin for carefully reading the manuscript.
The support from JCHP-FFE of the Forschungszentrum J\"ulich is
gratefully acknowledged.

\end{document}